# Measurements with Noise: Bayesian Optimization for Co-optimizing Noise and Property Discovery in Automated Experiments

Boris N. Slautin[1*], Yu Liu[2], Jan Dec[3], Vladimir V. Shvartsman[1], Doru C. Lupascu[1], Maxim Ziatdinov[4], Sergei V. Kalinin[2,4*]

[1] Institute for Materials Science and Center for Nanointegration Duisburg-Essen (CENIDE), University of Duisburg-Essen, Essen, 45141, Germany
[2] Department of Materials Science and Engineering, University of Tennessee, Knoxville, TN 37996, USA
[3.] Institute of Materials Science, University of Silesia, Katowice, PL 40-007, Poland
[3] Pacific Northwest National Laboratory, Richland, WA 99354, USA

We have developed a Bayesian optimization (BO) workflow that integrates intra-step noise optimization into automated experimental cycles. Traditional BO approaches in automated experiments focus on optimizing experimental trajectories but often overlook the impact of measurement noise on data quality and cost. Our proposed framework simultaneously optimizes both the target property and the associated measurement noise by introducing time as an additional input parameter, thereby balancing the signal-to-noise ratio and experimental duration. Two approaches are explored: a reward-driven noise optimization and a double-optimization acquisition function, both enhancing the efficiency of automated workflows by considering noise and cost within the optimization process. We validate our method through simulations and real-world experiments using Piezoresponse Force Microscopy (PFM), demonstrating the successful optimization of measurement duration and property exploration. Our approach offers a scalable solution for optimizing multiple variables in automated experimental workflows, improving data quality, and reducing resource expenditure in materials science and beyond.

# 1. Introduction

The rapid advancement of self-driven laboratories ranging from individual automated tools to labs integrating all stages from synthesis to characterization in a fully automated workflow, is currently transforming approaches in material exploration and experimental science as a whole.[1–5] One of the key drivers behind these groundbreaking changes is the development of advanced machine-learning (ML) algorithms that can make decisions previously fully reliant on humans[1,6–8]. The capability for decision-making is the key difference between simple automation with strict algorithms and higher-level automation used in scientific research. Simple automation, which has been successfully implemented in various industrial processes at least since Henry Ford's time, follows strict algorithms with predefined rules and procedures. In contrast, automation in exploration and optimization processes required for e.g. scientific research often involves iterative workflows with complex decision-making at each step.[5] This allows for dynamic adjustments of the experiment trajectory based on real-time data and optimization strategies. While the complete exclusion of the operator from the automated loop remains impossible, the human role is gradually changing for more and more high-level decisions and mind-related work. Meanwhile, routine experimental procedures, which rely heavily rely on real-time data analysis for decision-making, are increasingly being handled by automated systems [9–11].

Currently, the mainstay of automated experimentation (AE) workflows across numerous scientific and industrial fields are the sequential Bayesian optimization (BO) algorithms [12,13]. The process is conducted within a domain-specific experimental object space. BO-guided workflows have an iterative structure with an automated selection of the next candidate to be explored from the object space at each step. This selection is based on a predefined policy and reward function and is independent of direct human choice. The examples of BO-driven optimization could involve searching for an optimal composition within a phase diagram for material optimization,[14–16] adjusting parameters like pressure, temperature, laser fluence, etc. for pulsed laser deposition,[17] micro-drilling,[18] laser annealing,[19] or selecting locations in the image plane for spectroscopic measurements in microscopy,[20,21] and so on. The BO-based frameworks have proven to be effective in guiding experiments, ranging from single-modality explorations in combinatorial synthesis, scanning probe microscopy or electron microscopy to the co-orchestration of multiple modalities in one experimental cycle [14,15,22–24].

In many cases, BO methods are built on the Gaussian Processes (GPs) as surrogate models capable of interpolation and uncertainty prediction over the parameter space.[13] However, the standard GP methods are purely data-driven, limiting their efficiency in many situations.[25] Realistic physical systems are often associated with ample prior physical knowledge, such as laws and dependences. Incorporating this knowledge into the BO cycle can significantly enhance optimization efficacy.[26,27] The common way to incorporate physical knowledge lies in specifying the mean function or kernel (covariance function) of the GP or defining boundary conditions.[26,28] To embed more complex pre-acquired knowledge expressed in the form of high-dimensional datasets, advanced approaches such as Deep Kernel Learning can be utilized [21,29,30]. Recently it has been shown that the addition of the mean function as a probabilistic physical model in structural GP significantly increases the efficiency of exploration for materials synthesis,[31] combinatorial library exploration, and physics discovery.[25,32]

One of the key but seldom explored aspects of automated experiments is measurement noise. For purely data-driven applications, the noise level can be treated as a prior reflecting the degree of trust in the experimentally obtained observables, or as an optimization hyperparameter. However, for physical systems, noise is not just an abstract prior but a measurable quantity that depends on multiple factors, including experimental conditions and parameters. For instance, in imaging and spectroscopic techniques, the noise level might be influenced by exposure time, sensor sensitivity, environmental conditions, and other operational parameters. The noise level can be reduced by optimizing measurement parameters and by increasing the exposure time. However, increasing the duration of the experiment inescapably leads to the growth of the experiment times and hence costs. Thus, finding a trade-off between the cost of an experiment and the signal-to-noise ratio (SNR) is an optimization problem that needs to be solved in many experiments.

More generally, reducing the financial expense of experimentation lies among the ultimate goals of experiment automatization. Thus, the cost of the experiment is one of the primary criteria for estimating the efficiency of AE. In many cases, the financial expense of experimentation is directly proportional to the time it consumes. Globally, two realistic approaches to experimentation can be distinguished: 1) budget-constrained experiments – maximizing exploration or optimization efforts within a predefined budget limit (time) for the entire experiment; 2) unrestricted experiments, where an algorithm must determine the best possible experimental trajectory according to the real-time feedback and without strict time limitations. The time expenses are determined by the number of exploration steps and the duration of each iteration. Today, most BO approaches focus on optimizing the number of required exploration steps, with the duration of each step typically being predefined. More advanced multi-fidelity and orchestrated BO approaches navigate across different fidelities or modalities to balance the cost of each iteration with the potential outcomes, thereby enhancing the efficacy of exploration.[33–35] However, the cost and duration of steps for each modality are typically assumed to be known as a constant. Optimizing not only the number of steps (experimental trajectory) but also the duration (cost) of each iteration should advance the efficiency of AE. Such a budget-limited approach requires optimizing the entire experimental trajectory, allowing for variation in costs between different steps during the experiment. In contrast, in unrestricted experiments, we can optimize the cost of the individual step and use the predefined parameters for further investigation.

As a special case for automated experiments, we consider the intra-step optimization of the measurement times in the presence of noise. The exposure time is typically defined by an operator before the measurements and remains constant throughout the experiment. At the same time, for many spectroscopical methods (XRD, Raman, etc.) the exposure time is a major parameter that determines the signal-to-noise (SNR) ratio and therefore the amount of the gained information after each iteration. It is important to note that the significance of optimizing measured noise increases with longer exposure times. For example, in Raman spectroscopy of high-quality crystals, the accumulation time may not exceed 0.1 seconds. While, for Raman measurements of lithiated electrode materials, where conductivity is much higher, accumulation times can extend to minutes or even tens of minutes.[36] This extended duration makes noise optimization critical, as it can significantly impact the quality, reliability, and cost of the measurements. Precise measurements of scalar properties, such as the ultra-low DC electron conductivity of dielectric crystals, also follow this principle.

Here, we present a workflow for incorporating intra-step noise optimization into the BO GP cycle. The proposed workflow defines the optimal exposure time to balance the quality and cost of each iteration directly within the cycle of the automated investigation of the target property. We explore the two alternative approaches: a reward-driven noise optimization and a double-optimization acquisition function. We validate our method through simulations and real-world experiments using Piezoresponse Force Microscopy (PFM).

## 2. In-loop noise level optimization: a workflow

To introduce the noise optimization workflow, we consider a case of optimization of property $f(x)$ in 1D space $x$. For example, the $x$ might represent a compositional axis within a combinatorial library. Property $f$ can be either a scalar or a vector. Our goal is to simultaneously optimize the property $f$ and the noise associated with its experimental determination within a single optimization loop. To achieve this, we need to expand the input space of the optimization model by adding dimension—time ($t$). As a result, the optimization of property $f$ will be carried out in the ($x,t$) space. While $f(x)$ is independent of the measurement duration (exposure time), the noise ($Noise_f$) in measuring $f(x)$ is determined by the exposure time. In the general case, the noise level may also depend on the location along the $x$ axis. However, for simplicity, we assume the noise to be independent of $x$, such that $Noise_f(x,t) = Noise_f(t)$. As a result, we encounter an unusual situation where optimization of both $f(x)$ and $Noise_f(t)$ occurs in a 2D space, however, each of the variables depends on only one of the two dimensions of the input space. We also note that whereas $f(x)$ can be arbitrary, the measured noise is expected to be a monotonically decreasing function of measurement time and in many experimentally important cases follow simple behavior (e.g. for Gaussian white noise or 1/f noise).

At each iteration, the optimization process can be broken down into three sequential steps: 1) $GP$ modeling of $f$ and its uncertainty distribution within the input space ($x,t$), 2) construction of the acquisition function that incorporates the cost of the experiment for noise optimization, and 3) selection of the next location to discover at the extremum to the selected acquisition function in ($x,t$) space (Figure 1). While the third step remains similar to the classical BO approaches, the construction of the surrogate $GP$ model and the acquisition function are detailed below.

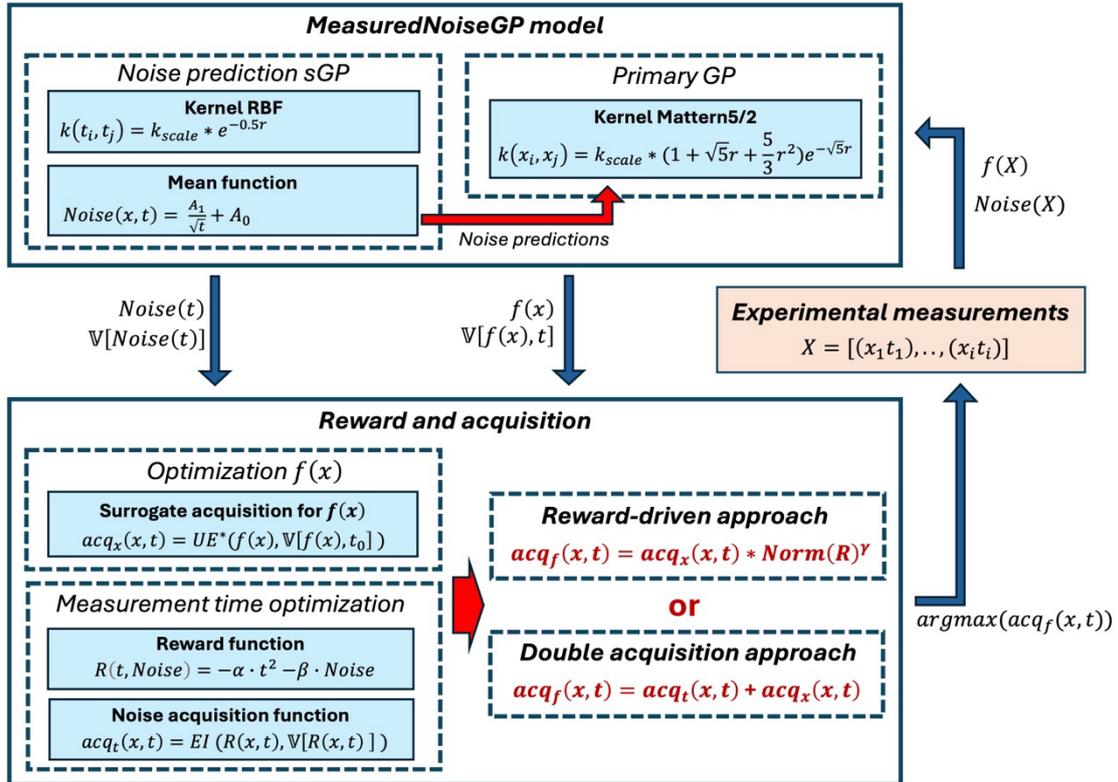

**Figure 1:** Scheme of double noise-property optimization workflow.

## 2.1 MeasuredNoiseGP for $f(x)$ predictions

The $f$ optimization is carried out in the low dimensional ($x,t$) space making reasonable direct implementing GPs as surrogate model for BO. Here we utilize the *MeasuredNoiseGP* model, which incorporates noise, estimated from the experimental observations, into the model rather than inferring it as in traditional $GP$.[37] Experimental noise in the measured points can be calculated directly from spectrum acquisition or derived from repeated measurements of scalar values. The noise level at unexplored locations is predicted independently of the $GP$ model used for $f$ (the *primary model*). This prediction is performed using a separate model (*noise model*). The noise model is trained to predict the noise component based on the noise estimated in the already measured locations. Once the noise predictions are obtained, they are added to the diagonal of the covariance matrix of the primary model. This adjustment accounts for the additional uncertainty introduced by the noise. The resulting amended covariance matrix ($K'$) for $f(x)$ in *MeasuredNoiseGP* model can be expressed as:

$$K' = K + diag(Noise_f), \qquad (1)$$

where $K$ is the covariance matrix for the primary model, and $Noise_f$ is the prediction of the noise according to the noise model. Hence, Eq. (1) enables reflection of the increased uncertainty due to the noise level.

To reflect the independence of the $f(x)$ on $t$, the kernel function (covariance) of the primary model has been adjusted by deactivating the time dimension. In this scenario the kernel function solely depends on the distance between the projections of the measured points onto the $x$ axis, ignoring their distribution relative to the $t$ axis. The specified Matérn 5/2 kernel is expressed as:

$$k(x_i, x_j) = k_{scale} * (1 + \sqrt{5}r + \frac{5}{3}r^2)e^{-\sqrt{5}r}, \qquad (2)$$

where $r = \left(\frac{x_i - x_j}{k_{length}}\right)^2$ is the squared distance normalized by the $k_{length}$. The $k_{scale}$ and $k_{length}$ are the kernel hyperparameters to be defined in $GP$ training. It is important to note that other $GP$ kernels can also be similarly adapted for this purpose.

In many cases, optimizing the noise level is facilitated by our understanding of its nature and the availability of models to describe it. The noise in the system typically consists of time-independent and time-dependent components. For many spectroscopical measurements, decaying noise arises as signal averaging is performed over longer acquisition times, with the noise level decreasing, following as $1/\sqrt{t}$. In turn, the time-independent noise component can be represented by a thermal, instrumental, and readout noise, for instance. The expression for the total noise may be written as:

$$Noise(t) = \frac{A_1}{\sqrt{t}} + A_0, \qquad (3)$$

where the constants $A_0$ and $A_1$ represent time-independent and time-dependent inclusions (Figure 2a). Given that the noise structure is described by expression (3), a structured $GP$ model with a specified mean function can be utilized to estimate the hyperparameters $A_0$ and $A_1$. It is crucial to ensure that the independence of noise from $x$ is accurately reflected in the noise model.

## 2.2 Acquisition function

The primary requirement for an acquisition function in BO is to identify the location that maximizes the expected benefit for exploration during the next iteration. This means that the acquisition function defined at the $(x, t)$ space should exhibit a maximum or minimum at the most promising location for further exploration. In our case of the in-cycle noise optimization, the constructed acquisition function should follow both policies 1) optimization/exploration of the target property $f(x)$ and 2) optimization of the exposure time.

It should be noted that, while the objective for noise investigation remains the same regardless of the experiment, the policy for property optimization is defined by an operator and may vary depending on the global experiment objective.

*2.2.1 $f(x)$ optimization*

Direct construction of traditional acquisition functions (such as Maximum Uncertainty, Expected Improvement, Upper Confidence Bound, etc.) based on $GP$ predictions and uncertainties over the $(x, t)$ space enable us to prioritize the location in $x$ axis for measuring $f(x)$ at the next step. At the same time, this approach is not suited for optimizing measurement duration.

To accomplish this goal, we proceed with the following derivation. From Eq. (1), the total uncertainty includes the time-dependent noise component and the time-independent part:

$$\mathbb{V}[f(x), t] = \mathbb{V}[f(x)] + Noise_f(t) \qquad (4)$$

The measurements of $f$ at each iteration introduce some additional knowledge and decrease the uncertainty. Concurrently, the noise level—dictated by our measurement capabilities and the exposure time—sets a theoretical limit, determining the maximum possible information gain and the minimum achievable uncertainty. In other words, we cannot define the $f(x)$ for some exposure time $t_0$ with lower uncertainty than $Noise_f(t_0)$. Lack of time dependence in the optimized property $f(x)$ means that projecting uncertainty onto the time axis will merely reflect the noise model behavior. Guided by classical acquisition functions and unable to effectively reduce uncertainty, the algorithm will typically select the maximum available exposure time. This occurs because, in such a direct approach, the cost function defining the optimal measurement time is not accounted for. In fact, the optimization of exposure time should be guided by a reward function, which incorporates experimental cost information into the acquisition function model.

*2.2.2 Noise reward function*

To optimize exposure time (and thereby reduce noise), it is necessary to introduce a cost model and a reward function ($R$) that balance the cost against the achievable noise level for a single measurement. In the simplest case, the cost of our measurements is determined solely by the exposure time and depends linearly on it – $R(Cost(t), Noise) = R(t, Noise)$

Given that our goal is to minimize noise while also minimizing cost (i.e., exposure time), our reward should decrease as either exposure time or noise increases. There are no strict rules to construct the reward function besides the principles mentioned above. In our experiments, we defined the reward function for noise optimization as follows:

$$R(t, Noise) = -\alpha \cdot t^2 - \beta \cdot Noise, \qquad (5)$$

where $\alpha$ and $\beta$ are coefficients. These coefficients are used to make the units of the noise and exposure time compatible and to balance their contributions within the reward function. We chose a quadratic dependence of the reward on the exposure time and a linear dependence on the noise level (Figure 2a). However, alternative combinations of these dependencies can also be considered. Given the distinct model $Noise_f(t)$, optimal exposure time can be defined from the profile of reward function along the noise model $R(t, Noise = Noise_f)$ (Figure 2b). To ensure that the reward function is well-scaled, we normalized it to the range [0,1].

Given the reward function, the acquisition function can be adjusted to capture the interdependency between the measurement cost (duration) and the noise level. Below, we outline two approaches for integrating noise optimization into the investigation of $f(x)$ by incorporating the reward function into the acquisition function:
1. **Pure Reward-Driven Noise Optimization**: In this approach, the reward serves as a weighting function for classical acquisition functions, which prioritizes certain exposure time. This method directly incorporates noise considerations into the reward mechanism.
2. **Double-Optimization Acquisition**: This approach uses an artificial total acquisition function that integrates two 'independent' components: the first is an acquisition function tailored for optimizing $f(x)$, and the second is a time acquisition function built around the noise reward function and the uncertainty in the noise model predictions. Each component addresses a distinct objective, enabling both noise and function optimization within a Bayesian framework.

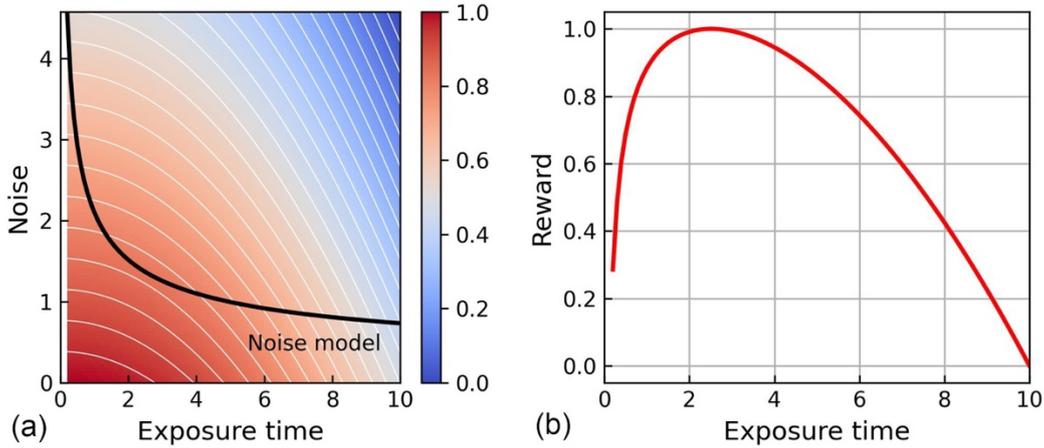

**Figure 2:** Reward function for exposure time optimization. (a) Distribution of the reward across the $(t, Noise)$ space. (b) Normalized profile of the reward function for the noise model, corresponding to the black line in (a). $A_0 = 0.1$, $A_1 = 2$, $\alpha = 0.1$, $\beta = 2$.

*2.2.3 Pure reward-driven approach*

In the pure reward-driven approach, the reward function acts as a weighting factor for the classical acquisition function defined over the entire $(x, t)$ space:

$$acq_f(x,t) = acq_x(x, t, \mathbb{V}[f(x), t]) * \text{Norm}(R)^\gamma \qquad (6)$$

where $acq_f(x,t)$ is the resulting acquisition function, $acq_x(x, t, \mathbb{V}[f(x), t])$ is the classical acquisition function determined by $x$, $t$ and full uncertainty $\mathbb{V}[f(x), t]$, and $\gamma$ is the exponent of the weighting function. The peak-shaped $R$ (Figure 2b) allows us to effectively prioritize the regions near the optimal time, while the acquisition function in areas corresponding to very short (high noise) and very long measurement durations (high cost) becomes smaller. The exponential factor $\gamma$ is used to control the steepness of the reward function (Figure 3a). The uncertainty of $f$ increases as $t$ approaches to zero, following a least $1/t$ (noise limit). The growth shifts the resulting acquisition function maximum when multiplied by the peak-shape reward function (Figure 3b). However, increasing the steepness of the reward function reduces this effect, mitigating the shift (Figure 3c).

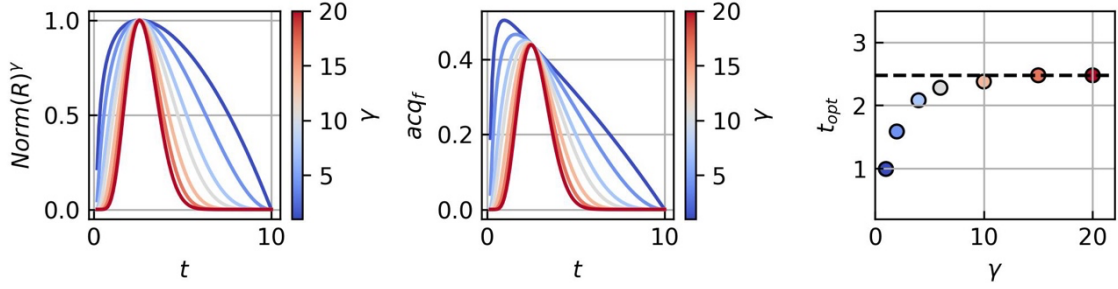

**Figure 3:** Reward-driven optimization: (a) Normalized reward functions (weighting functions) used to construct (b) acquisition functions for different exponential factors ($\gamma$). (c) Optimal time predictions as a function of the $\gamma$. The ground truth optimal time is represented by the black dashed line.

*2.2.4 Double-optimization acquisition function*

Optimizing exposure time within a Bayesian framework requires constructing an acquisition function that balances the reward with its associated uncertainty. Since the reward function is derived from the noise model, the uncertainty propagation rule can be used to calculate the uncertainty of the reward. Classical BO approaches with acquisition function (e.g., Expected Improvement, Upper Confidence Band, etc.) defined over $(x, t)$ space (nominally, because $R$ is independent on $x$) allow for the optimization of exposure time. Since we cannot reduce the $Noise_f(t_i)$ for the chosen point $(x_i, t_i)$, we construct a surrogate acquisition function, $acq_x(f(x), \mathbb{V}[f(x), t_0])$ to optimize $f(x)$ by extending the uncertainty of $f$ predicted at a specific time $\mathbb{V}[f(x), t = t_0]$ across the entire $(x, t)$ space.

The double-optimization acquisition function, $acq_f(x, t)$, should integrate both components described above: the exposure time optimization acquisition function and the selected acquisition function for optimizing $f(x)$. The final double-optimization acquisition function was constructed as a sum of the normalized acquisition functions for exposure time – $acq_t(x, t)$ and for $f(x) - acq_x(x, t)$.

$$acq_f(x, t) = \tfrac{1}{2}[Norm(acq_t(x, t)) + Norm(acq_x(x, t))] \quad (6)$$

## 3. Experimental

The efficiency of the method was evaluated both through simulations and in the actual exploration of the local domain structure using the Piezoresponse Force Microscopy (PFM) technique. The gpax Python library was used to implement the Gaussian Processes. To execute the workflow and principles outlined above, we modified both the $MeasuredNoiseGP$ model and some kernel functions within the gpax library. Within the modified optimization model, noise prediction was carried out using a structured $GP$ model with a specified mean function as in expression (3).

### 3.1. Theoretical validation

**In simulations** of the real experiment, we emulated the investigation of a scalar variable dependency, with the uncertainty in the value of the variable estimated from repeated measurements. The simulations of the actual exploration were performed using the Forrester function (Figure 4a).[38] Due to the complexity and non-linearity, the Forrester function is a widely recognized benchmark often used to assess the performance of optimization algorithms. It is defined as:

$$f(x) = (6x - 2)^2 \sin(12x - 4)), x \in [0,1] \quad (7)$$

The ground truth noise model is defined by equation (3) with parameters $A_0 = 0.1$ and $A_1 = 2$ (Figure 4b). The reward function for optimizing exposure time is given by expression (5) with $\alpha = 0.1$ and $\beta = 2$. The coefficients $\alpha$ and $\beta$ indirectly determine the cost function. The optimization process is constrained within the input space $x \in [0,1]$ and $t \in [0.2, 10]$.

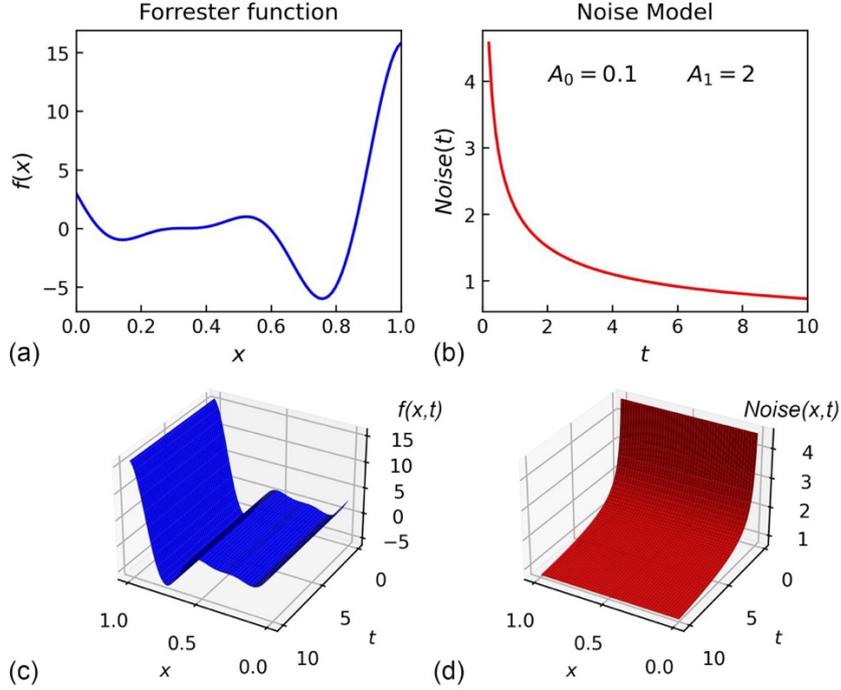

**Figure 4:** Ground truth (a) $f(x)$ function and (b) noise model; (c) $f(x)$ and (d) noise model defined over the exploration space.

In the simulations the $f(x)$ study was driven by the pure exploration strategy (Uncertainty Exploration, UE). The exponential factor $\gamma$ is set to 5. In the double-optimization approach, the EI acquisition function is used for the exposure time optimization. All simulation experiments start with 3 seed measurements in the random locations within input space and contain 15 subsequent exploration BO-guided steps. Before noise optimization converged, 30 repeated measurements of $f(x_i)$ were simulated at each iteration. These measurements are used to determine the function value and assess the associated uncertainty at each measured point $(x_i, t_i)$. The variance in the predicted time below $5 \times 10^{-4}$ for the last three steps is used as a criterion of the noise model convergence. After the model has converged, $f(x_i)$ is estimated using 3 measurements, while the noise level is determined based on the model predictions for $t_i$. The priors $A_0$ and $A_1$ in the noise mean function are sampled from a $Uniform(0, 15)$ distribution for both approaches.

### 3.2. Automated experiment

The **real automated experiment** evaluation of the model efficiency was performed using PFM using an MFP-3D (Oxford Instruments, USA) scanning probe microscope. Silicon probes with the conductive Pt/Cr coating Multi75 E-G are used for the response acquisition. The microscope is operated automatically via the AESPM Python library using a control program written in a Jupyter notebook.[39] The main calculations, including BO, are executed in Google Colab, which is connected to the microscope control notebook for real-time data exchange through a simple web server.

The experiment involved measuring the local electromechanical response to the AC voltage applied by the tip. This response arises from the converse piezoelectric effect, reflecting the dependence of the piezoelectric coefficient on the local domain structure. To estimate the dependence of local surface displacement on surrounding domain patterns, electromechanical resonances were measured across 180-degree domain structures in a high-quality PbTiO₃ single crystal oriented perpendicular to the [001] axis. The measured resonances are fitted using a simple harmonic oscillator (SHO) model:

$$r(\omega) = \frac{r_{real}\,\omega_0^2}{\sqrt{(\omega^2 - \omega_0^2)^2 + \left(\frac{\omega \omega_0}{Q}\right)^2}} \quad (8)$$

where $\omega$ is the frequency, $r(\omega)$ is the measured amplitude of the surface displacement, $r_{real}$ is the actual amplitude, $\omega_0$ the resonance frequency and $Q$ the quality factor. The uncertainty in the surface displacement, derived from the SHO fitting, was determined based on the approximate covariance of the fit. The resonance spectra were measured at a constant sampling rate of 10 kHz. Given the constant sampling rate, varying the sweep time led to changes in the number of acquired data points within the resonance curve, affecting the precision of the resonance fitting by Eq. (8) and, thereby, surface displacement amplitude estimation.

The objective of the real experiment was in the exploration of the piezoelectric response dependences (UE acquisition function) with simultaneous optimization of the sweep time (EI acquisition function). The double acquisition approach was used to guide the automated experiment. The experiment comprised 20 BO-guided steps, preceded by 3 preliminary seed measurements at random locations within the exploration space.

## 4. Experimental Results

### 4.1. Experiments Simulations

A simulation of the real automated experiment was employed as the first step of the workflow efficiency estimation. The experiment simulates the exploration of the Forrester function by the iterative measurements of $f(x)$ within the $x \in [0,1]$. At each exploration step, the automated agent selects the next location to be explored and the "exposure" time, which determines the precision of defining $f(x)$. The noise component is modeled as a normally distributed addition to the ground truth, with its standard deviation governed by the duration based on the ground truth noise model. Before the noise model converges, the algorithm estimates the noise level through repeated measurements at the exploration locations. Once the noise optimization converges, the noise estimation is derived from the optimized noise model. The idea and main goal of the experiment is to explore $f(x)$ automatically, simultaneously optimizing the measurement duration.

The experiment simulation, driven by both the pure reward (Figure 5a,c,e,g) and double acquisition approaches (Figure 5b,d,f,h), demonstrated their ability to converge after only a few BO-based exploration steps. However, notable differences in the evolution of the exploration process are observed between the two methods. We repeated the experiment in simulation mode multiple times for both approaches. The discussion below focuses on the comparison of the most representative results; additional outcomes are published on GitHub (see Data Availability).

In most of our experiments, even with only three seeding locations, the algorithm often succeeds in prioritizing duration ranges close to the true optimal time. However, the accuracy of this noise estimation heavily depends on the number of iterative measurements taken at each location. To ensure reliable estimation, we follow the empirical lower limit for parameter estimation of a normal distribution, using 30 measurements at each location.

The pure reward-driven approach demonstrates fast convergence after a few BO iteration steps (Figure 5c). Typically, the reward-driven model converges to a duration slightly lower than the ground truth value due to the combined effects of reward and noise, as discussed earlier (Figure 5a,c). Increasing the exponential factor, $\gamma$, further reduces this difference. It is important to note that, in the limit, the reward function can be substituted by a delta function with its peak at the optimal reward point. While the algorithm converges rapidly to the optimal exposure time, the estimation of the $A_0$ and $A_1$ parameters that define the noise model may not always align perfectly with the ground truth. The algorithm typically approximates the noise model accurately near the optimal exposure time, but predictions for both very high and very low durations may remain less accurate.

Experiments based on the double-acquisition method typically require a similar number of steps before the model converges to the optimal measurement duration (Figure 5d, Figure S1). Incorporating noise prediction uncertainty into the optimization process in a double-acquisition-driven approach typically improves noise model parameter predictions. However, this doesn't always lead to a significant enhancement in the accuracy of the optimal duration prediction. Additionally, the double-acquisition method does not exhibit the optimum shift effect, often resulting in a slightly more accurate estimation of the ground truth duration.

Concurrently with exposure time optimization, both algorithms explore the Forrester function using the UE acquisition function, which is the main objective of the automated experiment. Important to note, that although we selected the uncertainty exploration, any other acquisition functions without restrictions can be employed for the $f(x)$ exploration or optimization. The structure of the proposed approaches is capable of independent optimization of noise and function within the same cycle (Figure 5e-h). The time dependencies of the acquisition function at different locations exhibit similar profiles. Similarly, the projections onto the x-axis of time-constant profiles show analogous patterns. This is clearly illustrated by the visualization of the acquisition function values across the exploration space (Figure 5a,b).

In our experiments, both the pure reward-driven and double acquisition-based approaches achieved convergence to optimal duration in more than 90% of the attempts during the first 10 exploration steps independent of the seed locations. Beyond the simple exploration of the function $f(x)$, the algorithm was also tested for optimization using the EI acquisition function (Figure S2). In these experiments, we observe that the noise model usually converges at a similar rate (Figure S1). No clear correlation was observed between the rate of measurement duration optimization and the optimization of the $f(x)$ itself. This suggests that noise optimization does not interfere with the primary optimization process. It is important to note that both approaches successfully identify the optimal measurement duration with fewer than 30 repetitive measurements at each point (Figure S3), which can be crucial for real-world experimentation. However, as the number of measurements per point decreases, the number of optimization steps required for convergence naturally increases.

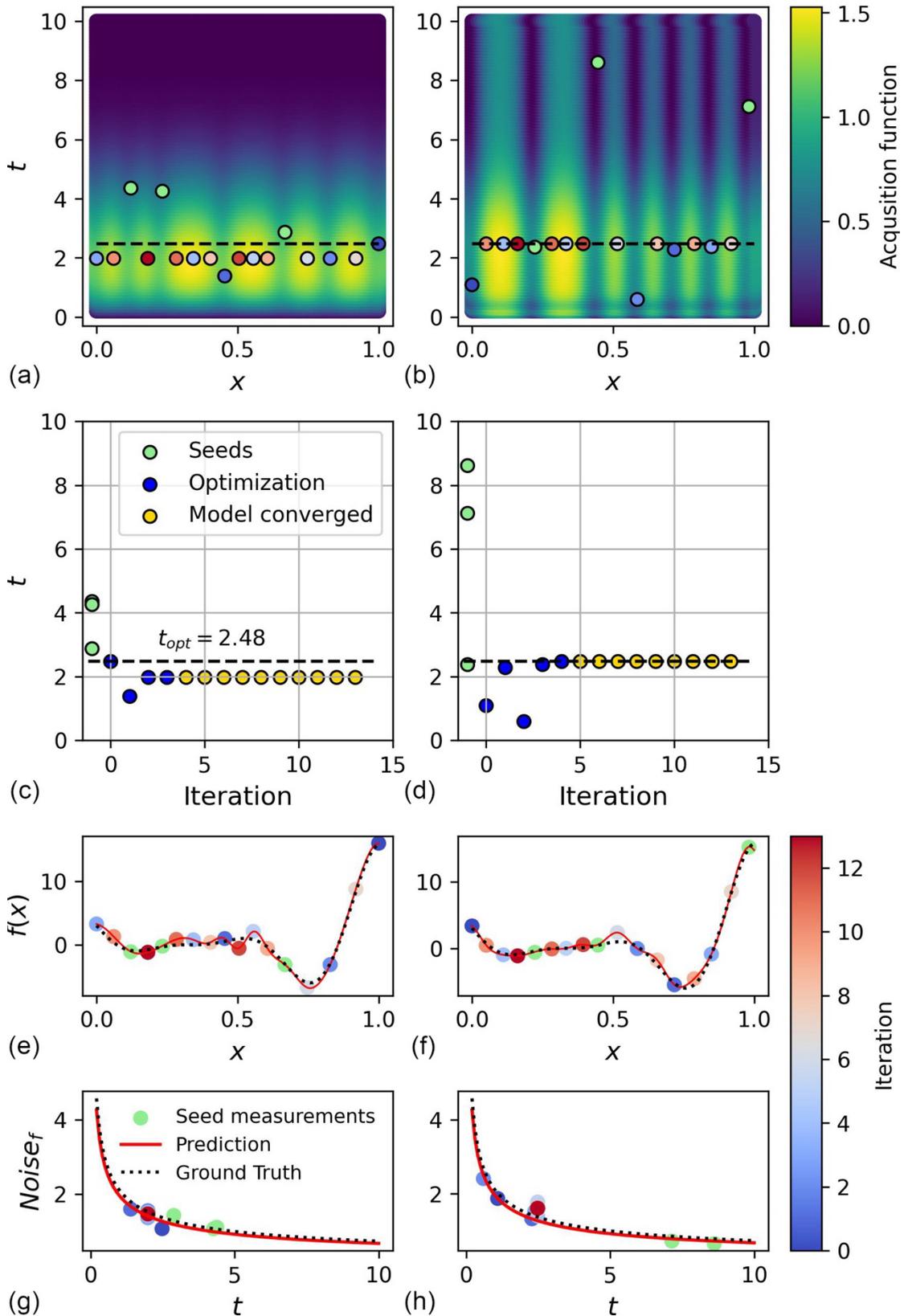

**Figure 5:** Automated experiment simulations using the (a, c, e, g) reward-driven approach and the (b, d, f, h) double-acquisition approach. (a, b) Experimental trajectories in the $(x, t)$ exploration space. The background shading represents the acquisition function values. (c, d) Evolution of the optimal measurement time predictions with iteration number. (e, f) Predictions of $f(x)$ and (g, h) $Noise_f(t)$ at the final exploration step. Data points in (a, b, e-h) are color-

coded based on iteration number, with light green indicating the initial seed measurements. The UE acquisition function drives both simulations.

*4.2. Real automated experiment*

We employed PFM contact resonance measurements to assess the algorithm efficiency in fully automated real experiment. The most precise approach for estimating local surface displacement induced by the piezoelectric effect involves fitting the contact resonance curve (Equation 8) obtained from the electromechanical response to an AC voltage frequency sweep applied by the SPM probe. The local surface deformation depends on the domain pattern in the vicinity of the probe. The primary objective of the experiment was to reconstruct a 35 µm-long profile of the electromechanical response as it traversed the ferroelectric domains in [001] cut of a high-quality $PbTiO_3$ single crystal (Figure 6a). The duration of each resonance measurement plays a crucial role in the accuracy of the response estimation and is subject to intra-step optimization. The time range considered for the spectrum measurement duration varied between 0.1 and 2 seconds. The $x$-axis in the study represents the distance from the starting point with the investigation profile.

The predicted response dependencies after 20 exploration steps, along with the actual data extracted from the PFM scan, are shown in Figure 6b. The predicted profile shape closely follows the real one measured by PFM, effectively reconstructing the actual profile. It is important to note that for accurate domain reconstruction, implementing structured GP as a surrogate model would be more appropriate than the GP with a modified Matérn 5/2 kernel used in this experiment. However, for our specific goals, the chosen model is more than adequate. The experiment trajectory clearly shows that, at each step, the algorithm selects spatial coordinates with the highest uncertainty, aligning with the UE acquisition function (Figure 6e). The close match between the predicted profile and the PFM scan response confirms the successful progression of the experiment toward the reconstruction objective.

From the analysis of the exploration trajectory (Figure 6d,e), it is evident that in the early stages, the model actively explores along the $t$-axis. However, after 11 initial exploration steps, the model converges to a value of single spectrum measurement duration 0.41 s (Figure 6c) representing the optimum for the chosen reward function (Equation 5). This convergence toward the optimal value after initial exploration is typical for BO driven by the EI acquisition function.

Overall, the proposed approach demonstrates its ability to pursue the predefined primary objective—an exploration of the PFM response along the profile—while simultaneously optimizing the resonance curve measurement duration. For greater statistical significance, we repeated our experiment multiple times with different seed locations. In each case, the model successfully converges. The results of these experiments can be found in GitHub (see Data Availability).

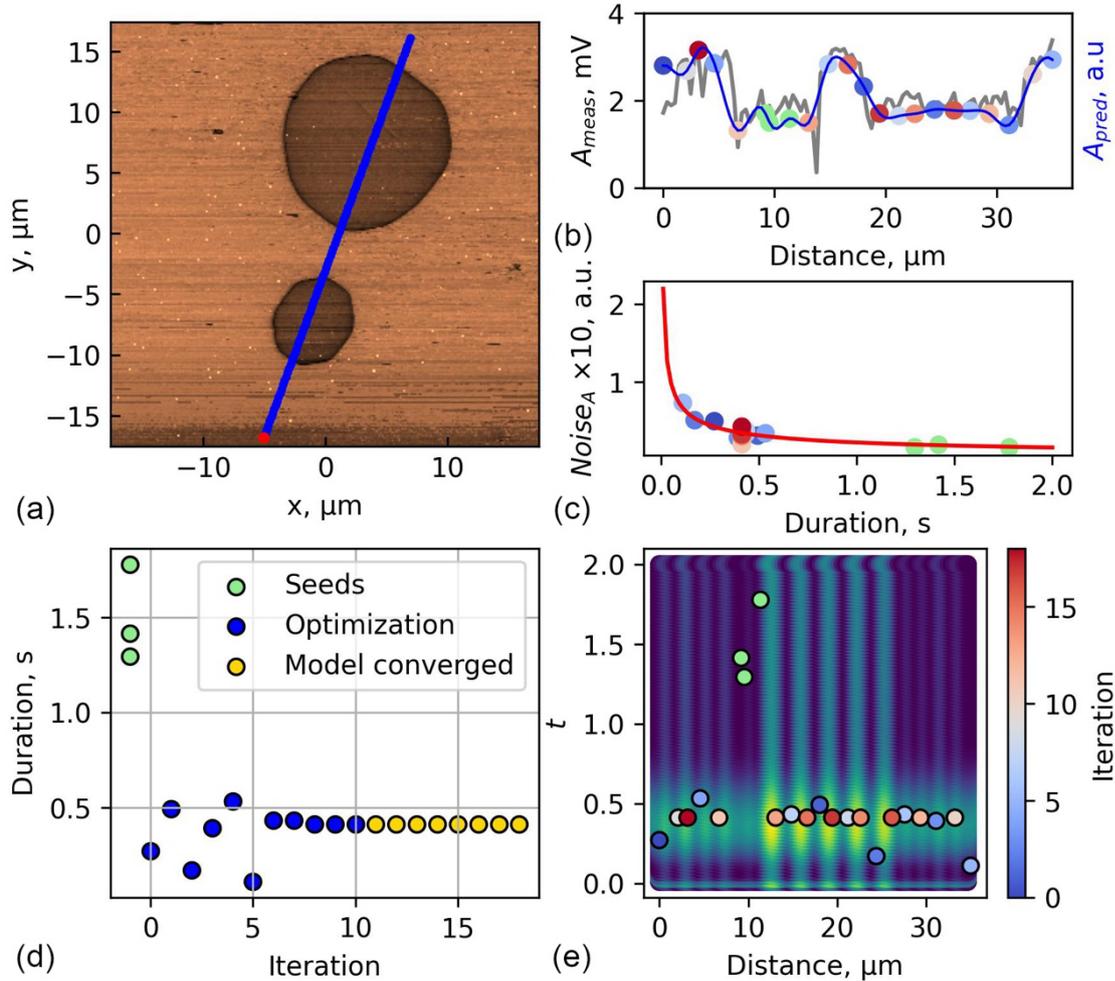

**Figure 6.** Results of the real automated experiment. (a) PFM amplitude scan with the exploration profile marked in blue. The red point corresponds to location 0. (b) Comparison of the predicted PFM amplitude profile with the actual data extracted from the PFM scan. (c) Prediction of the $Noise_A(t)$ along with the explored points. (d) Evolution of the predicted optimal measurement times as a function of iteration number. (e) Experimental trajectory visualized in the $(x, t)$ exploration space. The background shading represents the acquisition function values. Data points in (b, c, e) are color-coded based on iteration number, with light green indicating the initial seed measurements.

## 5. Summary

To summarize, we propose a workflow for optimizing measurement duration in real-time during automated experiments. Our algorithm employs the *MeasuredNoiseGP* model, modified with specialized kernels. Optimization within this workflow can be driven by either a pure reward-based approach or a double optimization acquisition function. The pure reward-driven approach relies solely on the prediction of the noise structure and demonstrates faster convergence to the optimal measurement duration. In contrast, the double optimization acquisition function, which incorporates noise model uncertainty into the process, offers a higher exploration impulse, leading to a more accurate reconstruction of the noise model, albeit requiring more iterations to converge.

The efficiency of the proposed framework was validated through a simulation of the Forrester function exploration, a benchmark for optimization processes, and a real automated PFM experiment. In both cases, the framework successfully demonstrates its ability to simultaneously optimize measurement duration and explore the target property. The proposed approach can be especially interesting for the automatization of spectroscopic measurements, such as Raman, XRD, etc., where the exposure (accumulation) time is one of the most important hyperparameters.

Although this workflow was designed for noise optimization, its principles can be extended to incorporate Bayesian optimization of multiple variables within a single experimental cycle. The proposed workflow introduces intra-step optimization within the automated experiment, which is expected to enhance global optimization efficiency by balancing knowledge acquisition with experimental costs.

**Data availability.**
The data that support the findings of this study are available in the supplementary material of this article. Code and raw experimental results are available without restrictions at [https://github.com/Slautin/2024_Noise_BO](https://github.com/Slautin/2024_Noise_BO). The GP codes are implemented using GPax package [https://github.com/ziatdinovmax/gpax.](https://github.com/ziatdinovmax/gpax.) The AESPM codes for the control of the microscope are implemented using AESPM package [https://github.com/RichardLiuCoding/aespm](https://github.com/RichardLiuCoding/aespm).

**Corresponding Authors**
Boris N. Slautin: boris.slautin@uni-due.de
Sergei V. Kalinin: sergei2@utk.edu

**Author Contributions**
**Boris N. Slautin**: Conceptualization (equal); Software (equal); Data curation; Writing – original draft. **Yu Liu**: Software (equal). **Jan Dec:** Resources; **Vladimir V. Shvartsman:** Writing – review & editing (equal). **Doru C. Lupascu:** Writing – review & editing (equal). **Maxim A. Ziatdinov**: Software (equal). **Sergei V. Kalinin**: Conceptualization (lead); Supervision; Writing – review & editing (lead).
All authors have given approval to the final version of the manuscript.

**Acknowledgment**
This research (workflow design, SVK) was primarily supported by the National Science Foundation Materials Research Science and Engineering Center program through the UT Knoxville Center for Advanced Materials and Manufacturing (DMR-2309083). VVS and DCL acknowledge the support by the German Research Foundation (DFG Project "Molectra" GR 4792/4-1, Project number 510095586). The development of the GPax Python package (MAZ) was supported by the Laboratory Directed Research and Development Program at Pacific Northwest National Laboratory, a multiprogram national laboratory operated by Battelle for the U.S. Department of Energy.

*Supplementary information*
**Measurements with Noise: Bayesian Optimization
for Co-optimizing Noise and Property Discovery in Automated Experiments**

Boris N. Slautin[1*], Yu Liu[2], Jan Dec[3], Vladimir V. Shvartsman[1], Doru C. Lupascu[1], Maxim Ziatdinov[4], Sergei V. Kalinin[2,4*]

[1] Institute for Materials Science and Center for Nanointegration Duisburg-Essen (CENIDE), University of Duisburg-Essen, Essen, 45141, Germany
[2] Department of Materials Science and Engineering, University of Tennessee, Knoxville, TN 37996, USA
[3.]Institute of Materials Science, University of Silesia, Katowice, PL 40-007, Poland
[3] Pacific Northwest National Laboratory, Richland, WA 99354, USA

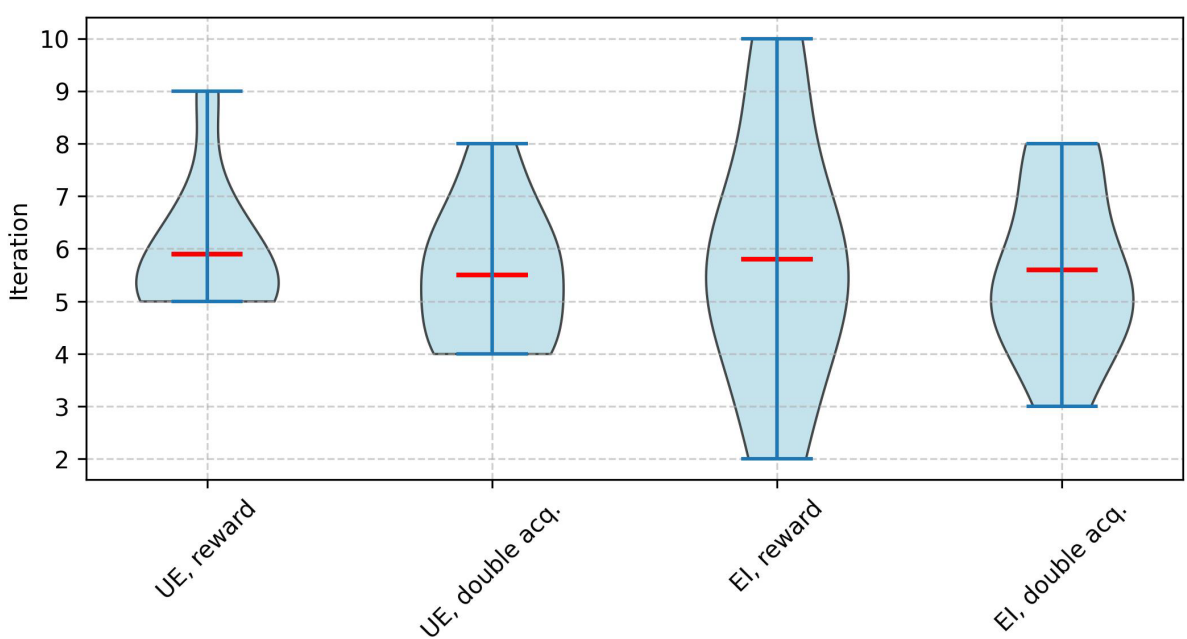

**Figure S1:** Violin plots showing the distribution of iteration numbers at which the noise model converges to the optimum. The simulations are driven by two acquisition functions (UE and EI) and different policies (pure reward-driven and double acquisition). Each violin represents data from 10 experiments, with red dashes indicating the mean values.

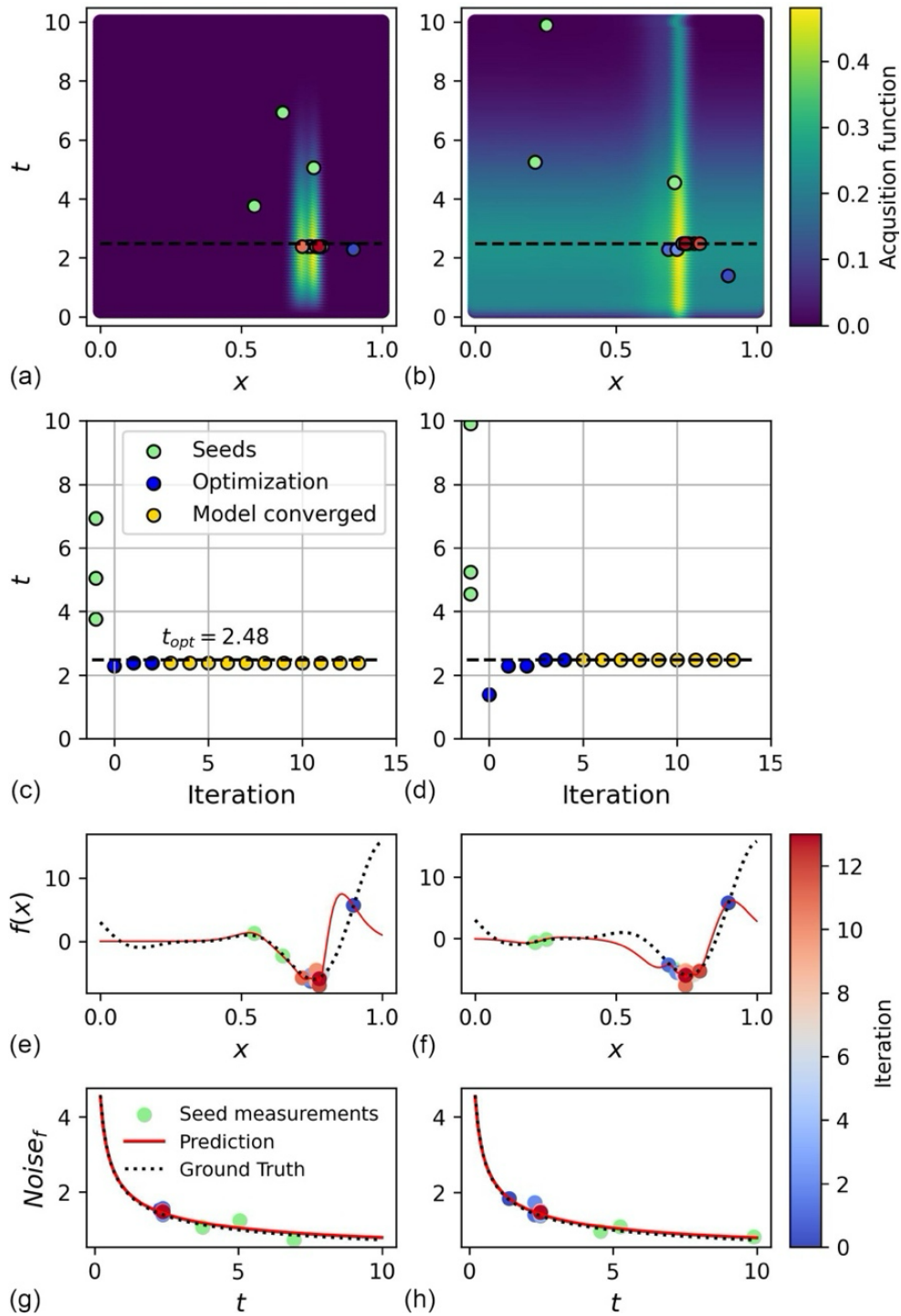

**Figure S2:** Automated experiment simulations using the (a, c, e, g) reward-driven approach and the (b, d, f, h) double-acquisition approach. (a, b) Experimental trajectories in the $(x, t)$ exploration space. The background shading represents the acquisition function values. (c, d) Evolution of the optimal measurement time predictions with iteration number. (e, f) Predictions of $f(x)$ and (g, h) $Noise_f(t)$ at the final exploration step. Data points in (a, b, e-h) are color-coded based on iteration number, with light green indicating the initial seed measurements. The **EI** acquisition function drives function optimization in both simulations.

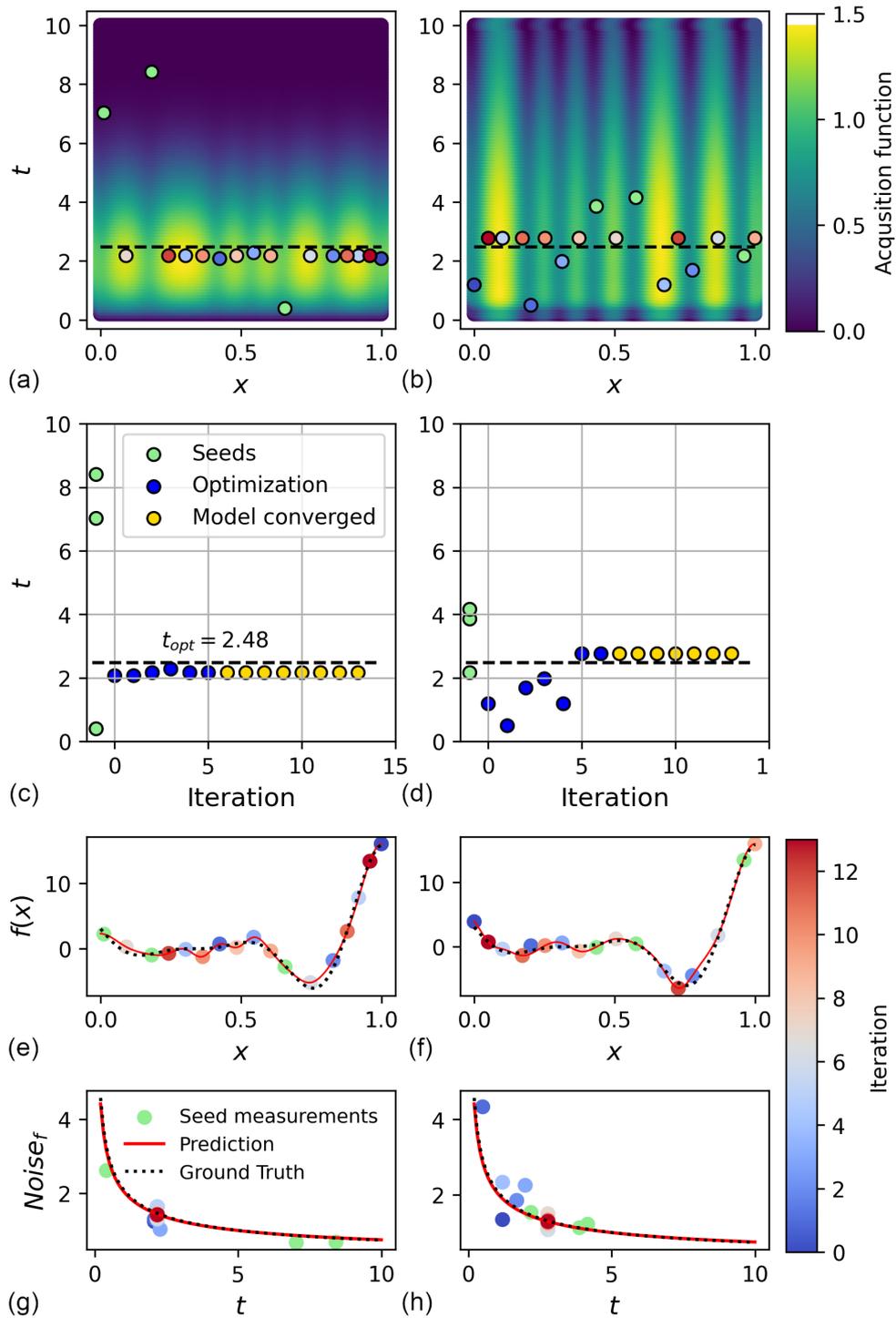

**Figure S3:** Automated experiment simulations using the (a, c, e, g) reward-driven approach and the (b, d, f, h) double-acquisition approach. (a, b) **Five** measurements are taken at each iteration to estimate both $f(x)$ and the $Noise$. Experimental trajectories in the $(x, t)$ exploration space. The background shading represents the acquisition function values. (c, d) Evolution of the optimal measurement time predictions with iteration number. (e, f) Predictions of $f(x)$ and (g, h) $Noise_f(t)$ at the final exploration step. Data points in (a, b, e-h) are color-coded based on iteration number, with light green indicating the initial seed measurements.